\documentclass{jpp}
\usepackage{graphicx}

\usepackage[utf8]{inputenc}
\usepackage[T1]{fontenc}
\usepackage{amsmath}
\shorttitle{Reconnecting plasmoid thruster}
\shortauthor{F. Ebrahimi}

\title{An Alfvenic reconnecting plasmoid thruster}

\author{Fatima Ebrahimi\aff{1}
  \corresp{\email{ebrahimi@pppl.gov}},
  }

\affiliation{\aff{1}Princeton Plasma Physics Laboratory, and the Department of Astrophysical Sciences, Princeton University, Princeton, NJ, 08544, USA}

\begin{document}

\maketitle
\begin{abstract}
   A new concept for generation of thrust for space propulsion is introduced. Energetic thrust is generated in the form of plasmoids (confined plasma
  in closed magnetic loops) when magnetic helicity (linked magnetic field lines) is injected into an annular channel. Using a novel configuration of static electric and magnetic fields, the concept utilizes a current-sheet instability to spontaneously and continuously create plasmoids
  via magnetic reconnection.
  The generated low-temperature plasma is simulated in a global annular geometry using the extended magnetohydrodynamic model.
Because the system-size plasmoid is an Alfvenic outflow from the reconnection site, its thrust is proportional to the square of the magnetic field strength and does not ideally
depend on the mass of the ion species of the plasma. Exhaust velocities in the range of 20 to 500~km/s, controllable by the coil currents, are observed in the simulations.

\end{abstract}
\section{Introduction}
Natural plasma engines such as the sun continuously generate enormous magnetic energy with complex field topology,
 and  release this magnetic energy in other forms. In the solar corona region, the linkage and the complexity of field lines, magnetic helicity, is injected through twisting field lines via shear motion of
their foot points. This build up of magnetic helicity is then released through the process of magnetic
reconnection, i.e. the rearrangement of magnetic field
topology of plasmas, in which magnetic energy is converted to kinetic energy and heat. On the surface of the sun, the process of
magnetic helicity injection provides the reconnection sites for oppositely-directed fields lines to come together to reconnect and energize.
In this letter, we introduce a novel thruster concept, which takes advantage of
a similar effect to convert magnetic energy to kinetic energy to 
produce thrust. In this concept, the reconnection sites are also generated via helicity injection, but by driving current along open field lines rather than twisting them via shear motion. This concept is based on the combination of \textit{two key physical effects, I) magnetic helicity injection and II) axisymmetric magnetic reconnection}. Significant thrust is generated in the form of plasmoids (confined plasma objects in closed magnetic loops) when helicity is injected into a cylindrical
vessel to induce magnetic reconnection. Existing space-proven plasma thrusters, including the ion thruster~\citep{Stuhlinger1964,Choueiri2009} and the Hall-effect thruster~\citep{Morozov1972,Raitses2007}, electrostatically accelerate ions to exhaust velocities $v_e$ of tens of km/s to produce thrust. However, for space exploration to Mars and beyond, high-thrust electromagnetic propulsion with exhaust velocities of tens to hundreds of km/s is needed. This new concept, capable of reaching high and variable exhaust velocities could complement existing designs for such missions.

For efficient propellant and propulsion-power use during space travel, thrusters should have an exhaust velocity similar to the velocity difference $\Delta{}v$ between the origin and destination celestial bodies.
This is quantitatively expressed by the Tsiolkovsky rocket equation,
\begin{equation}
\Delta{}v = v_e \ln{(m_0 / m_1)} \, ,
\label{eq:tsiolkovsky}
\end{equation}
where $m_0$ and $m_1$ are the total mass, including propellant, at the origin and destination, respectively. Eq.~\ref{eq:tsiolkovsky} shows that for a given $v_e$ and final mass $m_1$ a linear increase in $\Delta{}v$ requires an exponential increase in initial mass $m_0$. If the propellant is fully spent at the destination, the ratio $(m_0 - m_1) / m_0$ is the propellant mass ratio. For conventional chemical thrusters (rockets), the exhaust velocity is limited by the speed of chemical reactions to about 1-4~km/s (or specific impulse $I_{\mathrm{sp}}$ between 100 and 400 seconds, where $I_{\mathrm{sp}}=v_e/g_0$, where $g_0 = 9.8 m/s^2$ is the standard gravity). 
Conventional rockets are therefore efficient only for space missions that can be performed with a $\Delta{}v$ budget of about 4~km/s, e.g. a mission from low Earth orbit (LEO) to low Moon orbit. Even for a highly optimized mission from LEO to Mars, lasting 3-5 months and with a brief launch window every 2-3 years,
a $\Delta{}v$ = 6~km/s is needed. With an optimistic assumption of $v_e$ = 4~km/s, Eq.~\ref{eq:tsiolkovsky} gives a propellant mass ratio of 78\%, i.e. on launch from LEO more than three quarters of the mass is propellant. Thus only Earth's immediate neighbors in our solar system are within reach of conventional rockets.

To surpass the exhaust velocity allowed by limited chemical energy density and reaction rates, electromagnetic propulsion can be used ~\citep{goebel2008,dale2020future}. Existing space-proven plasma thrusters can reach a specific impulse $I_{\mathrm{sp}}$ of about a couple of thousands seconds (i. e. $v_e$ of about tens of km/s).
High-thrust electromagnetic propulsion with $I_{\mathrm{sp}}$ of tens of thousand of seconds is needed to explore the solar system beyond the Moon and Mars, as well as to rendevouz with asteroids, to deflect them if they are on a collision course with Earth, or to capture them for use as a source of water and construction materials to support human presence in space. The unique feature of the plasmoid thruster introduced here is its high and variable $I_{\mathrm{sp}}$, in the range 1,000 to 50,000 seconds, which would be a key advantage for space missions with a large $\Delta{}v$, i.e.
to Mars and beyond. Here, we show that these high specific impulses could be achieved through continuous production of plasmoids to accelerate ions via a magnetic reconnection process.

Magnetic reconnection, which is ubiquitous in natural plasmas, energizes many astrophysical settings throughout our solar system including corona (solar flares), solar wind, planetary interiors and magnetospheres [see~\citep{ji2020major} and references therein], as well as throughout our universe, such as flares from accretion disks around supermassive black holes~\citep{ripperda2020magnetic}. Magnetic reconnection causes particle  acceleration to high energies, heating, energy and momentum transport,~\citep{ebrahimitearing} and self-organization. The Parker Solar Probe~\citep{fox2016PSP} also provides access to a new frontier for exploring and providing observational evidence of large-and small scale reconnecting structures in the solar corona. In laboratory fusion plasmas plasmoid mediated reconnection has shown to be important during plasma startup formation~\citep{ebrahimi2015plasmoids}, nonlinear growth of an internal kink mode,~\citep{biskamp86,gunter15} as well as transient explosive events such as edge localized modes in tokamaks~\citep{ebrahimi2017ELM}. Here, we  demonstrate a practical application of plasmoid mediated reconnection, namely for space propulsion. 

The new type of plasma thruster we are here proposing uses an innovative magnetic configuration to inject magnetic helicity using two annular electrodes biased by a voltage source, thereby inducing spontaneous reconnection via formation of a current sheet, which continuously breaks and generates plasmoids.
The concept of 
biasing open field lines to stretch lines of force and form "plasma rings"~\citep{alfven60} was first introduced 
in the so-called coaxial plasma gun (accelerator) experiments in 1960~\citep{alfven60,marshall1960}. Since then, coaxial (annular) plasma accelerators have been extensively used and evolved for various applications,
including for fusion plasmas to form spheromaks~\citep{jarboe83,hsu2003,mclean2002} and to fuel tokamaks with compact toroids~\citep{brown90,ring_CT88,raman_CT94}. The plasma accelerator has also been proposed as a magnetoplasmadynamic (MPD) thruster for propulsion applications~\citep{Schoenberg93,cheng_gun71} and for generating high-velocity plasma jets~\citep{witherspoon2009}.
In all these annular plasma accelerators the Lorentz $\textbf J \times \textbf B$ force 
generated by a self-induced magnetic field accelerates plasmas to large velocities. In our new concept the acceleration is instead due to magnetic reconnection~\citep{zweibel09,yoo2017electron}.  Unlike existing plasma accelerators, the thrust is generated from the acceleration of
bulk fluid due to continuous formation of reconnecting plasmoids in the magnetohydrodynamic (MHD) regime. Neither external pulsing nor rotating fields ~\citep{bathgate2018thruster,2017JPlPh} are required here for acceleration through reconnection.

Axisymmetric reconnecting plasmoids are secondary magnetic islands, which are formed due to plasmoid instability. At high Lundquist number,  the elongated current sheet becomes MHD unstable due to the plasmoid
instability~\citep{biskamp86,tajima_shibata97,loureiro07,daugthon09,ebrahimi2015plasmoids,2016luca}, an example of spontaneous reconnection. The transition to plasmoid instability was shown to occur when the local Lundquist number $S = L V_A/\eta$ ($V_A$ is the Alfven velocity based on the poloidal reconnecting magnetic field, L is the current sheet length, and $\eta$ is the magnetic diffusivity) exceeds a critical value (typically a few thousand). Our thruster
concept is based on the formation of this elongated current sheet for triggering fast reconnection and plasmoid formation. Effects beyond MHD may also contribute to fast reconnection as the current sheet width ($\delta_{\mathrm{sp}}$) becomes smaller than the
two-fluid or kinetic scales~\citep{cassak2005,ji2011}. However, for thruster application we desire system-size MHD plasmoid formation (with radius ranging from a few to tens of centimeters), where kinetic effects become subdominant for low-temperature plasma (in the range of a few eV to a couple of tens of eV). Here, the MHD plasmoid mediated reconnection  occurs at
high Lundquist number (about $10^4$ and above), which is achieved
at high magnetic field rather than low magnetic diffusivity (or high temperature).
To  form a single or multiple X-point reconnection site, oppositely-directed biased magnetic field (in the range of  20-1000G) is injected through a narrow gap in an annular device.  We find that the plasmoid structures demonstrated in resistive (or extended)  MHD simulations produce high exhaust velocity and thrust that scale favorably with applied magnetic field. It will be shown that the fluid-like magnetic plasmoid loops continuously depart the magnetic configuration about every 10 $\mu s$  with Alfvenic velocities in the range of 20 to 500 km/s, and the thrust does not ideally depend on the mass of the ion species of the plasma.

\section{ Schematics of the thruster}
\begin{figure}
\centering
  \includegraphics[width=3.8in,height=3.8in]{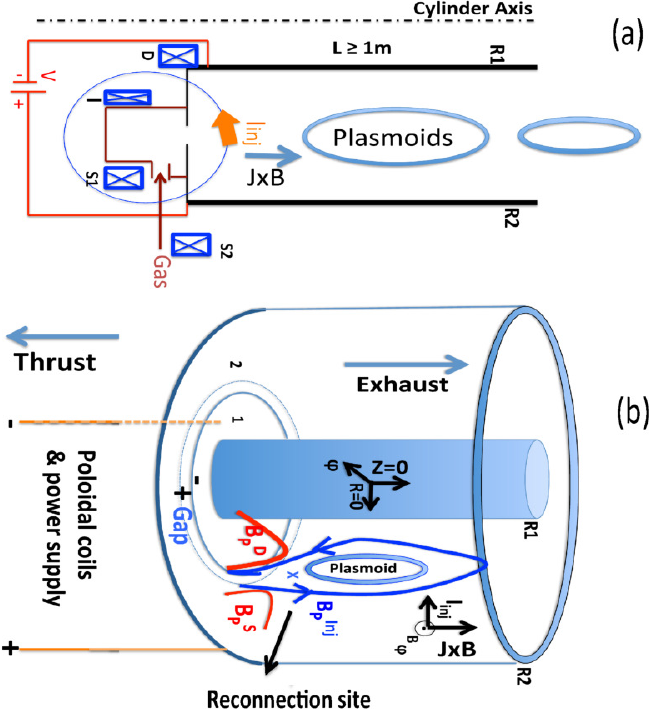}
\caption{A schematic of (a) the vertical cross-section and (b) the entire domain of the reconnecting plasmoid thruster.
In an annular configuration, injected poloidal field $B^{\mathrm{inj}}_P$ (blue circle) is generated by poloidal field injector coil (I), while  current ($I_{\mathrm{inj}}$) is pulled along open field lines by applying $V_{\mathrm{inj}}$. Numbers 1 and 2 show inner and outer injector biased disk plates, respectively, separated by the injector gap. All the axisymmetric poloidal coils (I, D, S1, S2) are located to the left of these plates. For formation of an elongated current sheet to induce spontaneous reconnection, the detachment coil D and shaping coils S1 and S2 are also energized to generate the poloidal fields $B^D_P$ and $B^S_P$, (shown in red in (b)).}
\label{fig:fig1}
\end{figure}
Figure~\ref{fig:fig1} shows the main parts of the reconnecting plasmoid thruster in an annular configuration.
Magnetic-helicity injection starts with an initial injector poloidal field ($B^{\mathrm{inj}}_P$, in blue, with radial, R, and vertical, Z, components), connecting the inner and outer biased plates in the injector region. Gas is injected and partially ionized by applying an injector voltage $V_{\mathrm{inj}}$ of a few hundred volts between the inner and outer plates (indicated by numbers 1 and 2), which also drives a current $I_{\mathrm{inj}}$ along the open magnetic field lines.
Plasma and
open field lines expand into the vessel when the Lorentz
force $J_{\mathrm{pol}} \times B_{\phi}$ exceeds the field line tension of the injector 
poloidal field. The azimuthal ($\phi$) field shown here, $B_{\phi}$, is  generated through injector current ($I_{\mathrm{inj}}$) alone (by applying $V_{\mathrm{inj}}$), or can be provided externally.
The plasma formation through electron impact ionization has been widely used by plasma accelerators and other helicity injection experiments~\citep{jarboe83,hsu2003,mclean2002,brown90,ring_CT88,raman_CT94,witherspoon2009,raman2003,raman2011experimental}. The conventional Townsend avalanche breakdown theory is applicable for coaxial helicity injection experiments~\citep{hammond2017,private}, a configuration similar to the thruster proposed here.

Up to this point the concept of magnetic helicity injection through the linkage of the injected poloidal field and injected azimuthal field from poloidal current along the open field lines 
is similar to the conventional annular accelerators.
However, at this stage we introduce the new concept of plasmoid-mediated reconnection for generating thrust, i.e. through forming a vertically elongated (along z) azimuthal current sheet ($J_{\phi}$),
which contributes to the Lorentz force.
To continuesly form a current sheet at the reconnection site, the detachment and shaping poloidal fields, $B_P^D$ and $B_P^S$ (shown in Fig. 1(b) and produced by the D, S1 and  S2 coils) are utilized and have an instrumental role for this thruster concept. These coils can be effectively used to strongly and radially squeeze the injector poloidal field  to cause oppositely directed field lines in the Z direction (shown in blue
arrows at the reconnection site) to reconnect. To form this reconnection site, the currents in the detachment and shaping coils are in the opposite direction of the current in the injector coil, and the detachment-coil current is of equal or larger magnitude than the injector-coil current. As a result, azimuthally symmetric system-sized plasmoid structures are detached and ejected to produce thrust.

\section{Global extended MHD simulations}
\begin{figure}
\centering
  \includegraphics[width=3.5in,height=3.2in]{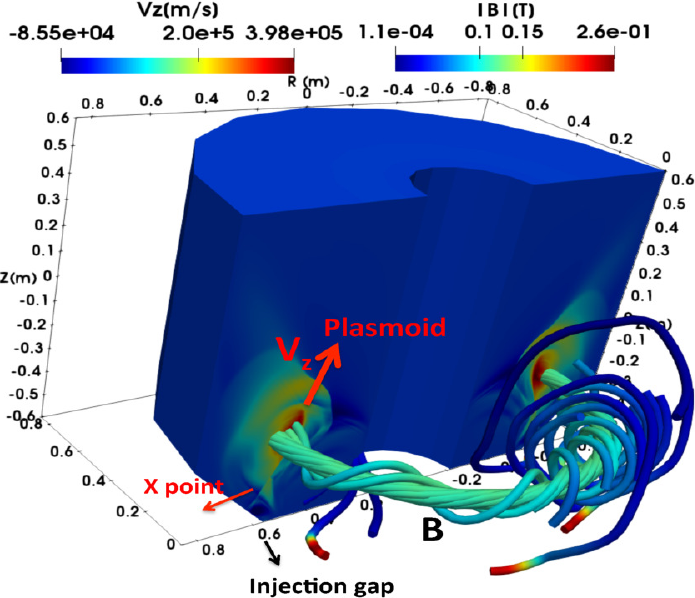}
  \caption{The formation of momentum carrying plasmoid during 3-D global extended (two-fluid) MHD simulations. The computational domain and the poloidal coil configurations are the same as the schematics in Fig 1. Plasmoid  ion (helium) velocity $V_z$ is seen in the poloidal (R-Z) cross section. The velocity structure remains azimuthally symmetric.  Following of the magnetic field line shows a closed magnetic loop associated with the plasmoid formation during reconnection.}
  \label{fig:fig2} 
\end{figure}

\begin{figure}
\centering
  \includegraphics[]{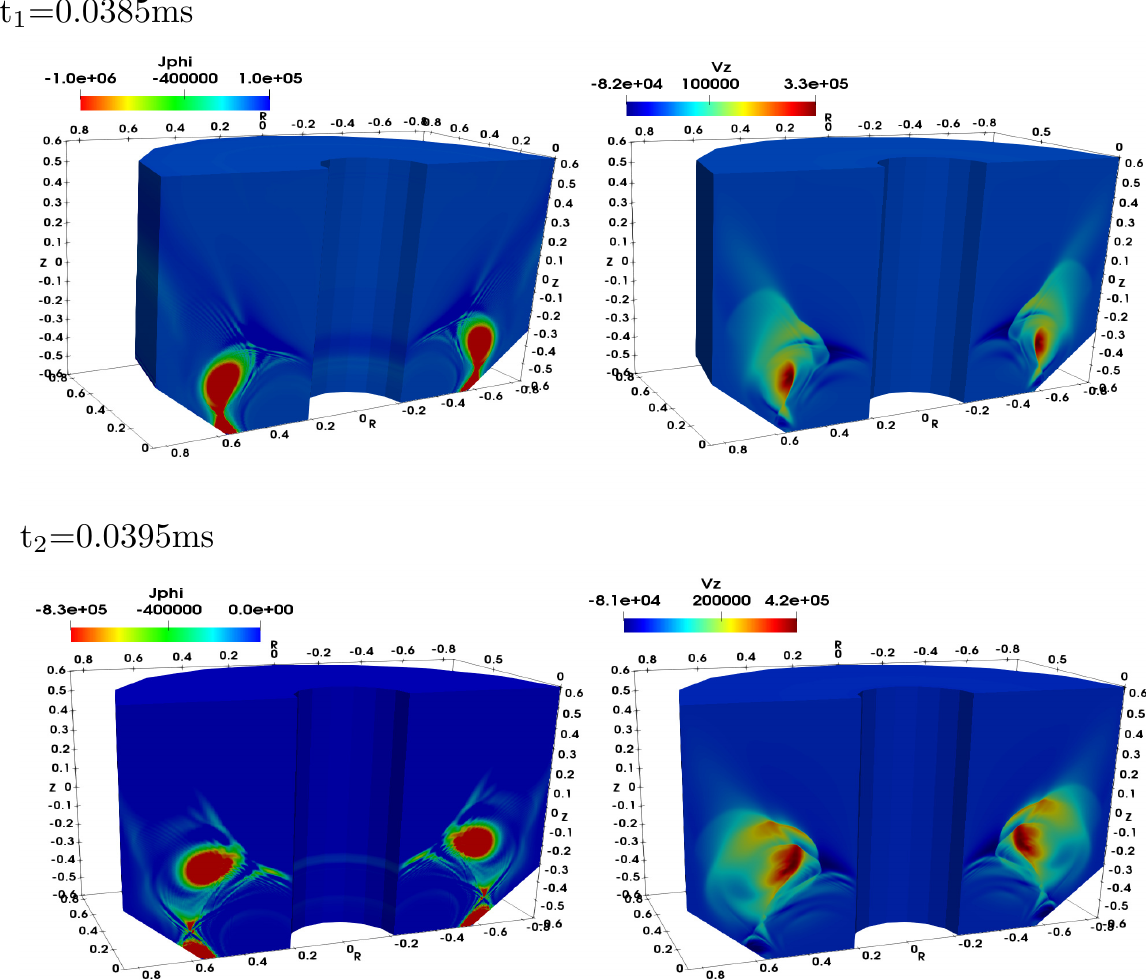}
  \caption{Current density [A/$m^2$] and the axial velocity[m/s] for the same two-fluid simulation shown in Fig. 2, at the two times $\mathrm{t_1}$= 0.0385 and at 1$\mu$s later at $\mathrm{t_2}$=0.0395.}
  \label{fig:fig2_2} 
\end{figure}

We  perform time-dependent extended MHD simulations of the thruster using the NIMROD code~\citep{sovinec04}, which is
a community code supported by DOE, and has been extensively used and validated for various helicity injection fusion experiments~\citep{hooper2012,adamchi,morgan2019formation}, including startup helicity injection experiments for spherical tokamaks (NSTX and NSTX-U)~\citep{ebrahimi2013,hooper2013,ebrahimi16}. We model coil currents that produce the needed injected field for the reconnection site.
Simulations are performed for a constant-temperature
model (pressure is not evolved in time) with constant-in-time poloidal-field coil currents.  We have optimized and varied extensively the current in the poloidal coils (I,D, S1,2) to form a
reconnection site and a current sheet.
Our extended MHD model consists of combined Faraday and generalized Ohm's laws and the momentum equation, \begin{equation}
\rho (\frac {\partial \textbf V }{ \partial t } 
 + \textbf V . \nabla \textbf V) = \textbf J \times
 \textbf B - \nabla . \Pi
\end{equation}
\vspace{-4mm}
\begin{equation}
 \frac {\partial \textbf B }{ \partial t } =
- \nabla \times ( -\textbf V\times \textbf B + \eta \textbf J + \frac{1}{ne}
\textbf J\times \textbf B + \frac{m_e}{ne^2}\frac{\partial \textbf J}{ \partial t})
\end{equation}
where \textbf V is the center-of-mass velocity and $\rho$ is the mass density
of a plasma with magnetic field \textbf B, and current density \textbf J,  The stress tensor ($\Pi$), which is treated as 
$-\rho \nu \nabla^2 \textbf V$ or $-\rho \nu W$, where $\nu$ is the 
kinematic viscosity and W is the rate of strain tensor. In all simulations, the kinematic viscosity is chosen to give Prandtl number Pm = $\eta/\nu$  = 2 - 7.5.
The magnetic diffusivities used in the simulations ($\eta=8-32 m^2/s$) are equivalent to constant low temperatures of $T_e \approx 5 - 14 eV$,
according to the Spitzer resistivity relation ($\eta [m^2 s^{-1}] = 410 \, T^{-3/2} [eV]$).
A constant electric field 
is applied across the narrow injector gap (located between the two injector plates R=0.54--0.57m, shown in Fig.~2). Perfectly-conducting boundary conditions
with no slip are used, except at the injector gap, which has a
normal $E\times B$ 
flow, where a constant-in-time electric field
is applied.~\citep{ebrahimi2013,hooper2013,ebrahimi2016dynamo} We use a poloidal grid with 45 $\times$
90  sixth-order finite elements in a global (R,Z) geometry and azimuthal mode numbers ($n_{\phi}$) up to 22 modes. A uniform number density (n) of 4 $\times10^{18}m^{-3}$ for a deuterium or helium plasma is used.
Simulations are performed with various coil currents in a straight plasma domain configuration, shown in Fig.~1, in a thruster channel with inner and outer radii at $R_1$=0.21m and $R_2$=0.85m, with the injector plates (1 and 2, shown in Fig. 1) located at Z=-0.8m. In general the results does not vary with the axial length of the thruster within the range 1-2m investigated, and with the angle of the lower injector plates  with the side walls (90$^{\circ}$ and 145$^{\circ}$).
The locations of the coils are adjusted for simulations performed in different domain sizes. In the simulations shown in Figs. 3 and 4, the injector (I), detachment (D), and shaping field (S1, S2) coils are located at R = 0.52, 0.31, 0.76, 1m and Z = -1, -0.82, -1.05, -0.81m, respectively. 
Fields $B_P^D$ and $B_P^S$ are static and assumed to have penetrated through the bounding surfaces.

A cut of the general computational domain of annular geometry is shown in Fig. 2. In this simulation we have used a two-fluid (2fl) model, i. e. using both the Hall term and electron
 inertia terms in Eq. 2 with an increased electron mass ($m_i/m_e$=73). The exhaust vertical velocity  of azimuthally symmetric plasmoids reaches about 400 km/s (Fig.~\ref{fig:fig2}). Most of the momentum is along the large plasmoid (and current sheet). The Alfvenic-type outflows obtained here ($V_A$ for this case using $B_z$=0.08 is about 460 km/s) is due to spontaneous reconnection and the plasmoid ejection. In the two-fluid model fast reconnection is caused by the Hall current,
which is a signature of the decoupling of electron and ion motion
at scales below the ion skin depth $d_i$ ($d_i=c/\omega_{pi}$, where
c is the speed of light and $\omega_{pi}$ is the ion plasma frequency). Here $d_i$ is calculated to be relatively large about, 11 cm. Current density and the axial velocity for the two-fluid simulation at two times are also shown in Fig.~\ref{fig:fig2_2}. The formation of the single X-point during Hall reconnection at $\mathrm{t_1}$ and then the subsequent ejection of a large plasmoid at $\mathrm{t_2}$ with large exhaust velocity of 420km/s are observed. The two-fluid results in terms of exhaust speed of plasmoids during reconnection is similar to the resistive MHD model (shown below). We therefore below present more detailed results from resistive MHD simulations to focus on the plasmoid mediated reconnection for this concept.
\begin{figure}
\centering
      \includegraphics[width=4.5in,height=2.9in]{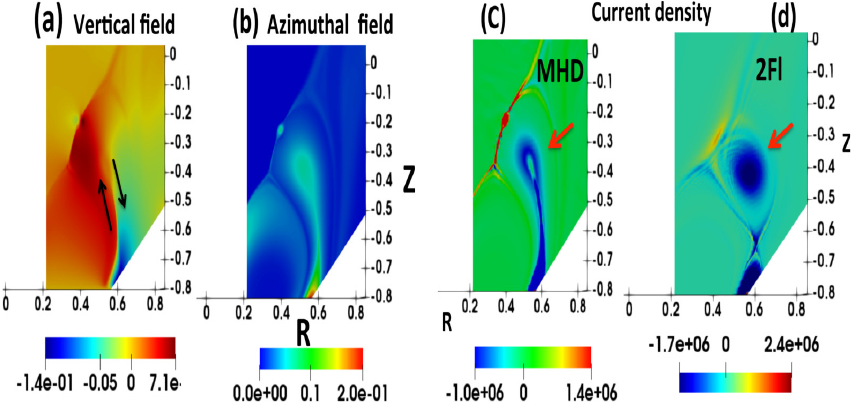}
\caption{R-Z cut of (a) injected vertical magnetic field $B_z$[T] (b) generated azimuthal field $B_{\phi}$[T] at t=0.044ms during the helicity injection (black arrows show the oppositely directed reconnecting $B_z$ field at the reconnection site). Azimuthal current density $J_{\phi}[A/m^2]$ with (c) MHD and (d) 2-fluid model (from same simulation as results in Fig. 2). Red arrows show reconnecting plasmoids.}
\label{fig:fig3}
\end{figure}
 
For this thruster concept, we start with an initial
injector 
poloidal field ($B^{\mathrm{inj}}_P$) with a very narrow footprint (where open field lines intersect the inner and outer plates) to form a reconnection site.  Poloidal R-Z cut of the oppositely directed injected reconnecting field ($B_z$), which provides the primary reconnection site, and the azimuthal field, which is intrinsically generated by the poloidal injector current in the injection region are shown in Fig.~\ref{fig:fig3}(a-b), respectively.
As the injector voltage is applied (by ramping $V_{\mathrm{inj}}$ from zero to about 200 V), the generated azimuthal field could reach as high as 2000 G (Fig.~\ref{fig:fig3}(b)), which causes the injector poloidal field to start expanding in the thruster channel. As the field is expanded in the domain,  the static fields  $B_P^D$ and $B_P^S$ radially pinches the injector field around the injector gap to form a primary exhaust reconnecting current sheet, as shown in blue (Fig.~\ref{fig:fig3}(c)).
The plasmoid instability is here triggered at local Lundquist number S $\sim$ 12,000 (based on $B_z \sim 500G, L = 0.5m$, and $\eta=16 m^2/s$).
The formation of a plasmoid along the current sheet is seen in Fig.~\ref{fig:fig3}(d) (in blue). For comparison,  the exhaust current density  for the two-fluid simulations (shown in Fig. 2) with single X-point Hall reconnection topology~\citep{huang2011hall} is shown in Fig.~\ref{fig:fig3}(d).  We have also performed 2-fluid simulations at smaller $d_i$ (of about 2 cm at higher density), and then recover the elongated single-fluid current sheet.

\begin{figure}
\centering
   \includegraphics[width=4.6in,height=2.5in]{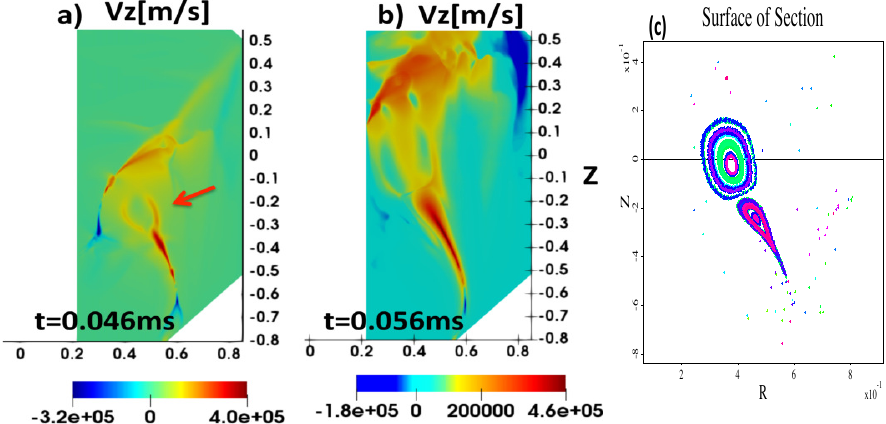}
\caption{Poloidal (R-Z) cuts of vertical flow velocity, (a) at t=0.046 ms, (b) t=0.056ms, (c) Poincare plot. Plasmoid ejection is shown by red arrow. }
\label{fig:fig4}
\end{figure}

The formation of a large plasmoid with exhaust outflow that reaches as high as $V_z$ = 400km/s around the reconnection region at later time t=0.046ms is shown in Fig.~\ref{fig:fig4}.
The maximum exhaust velocity is mainly along the current sheet and the plasmoid(as can be seen in Fig.~\ref{fig:fig4} a-b) and exceeds the visco-resistive outflows of the MHD Sweet-Parker (S-P) model, 
$V_{out}/V_{A(\mathrm{pol})} = 1/\sqrt{1+Pm} \sim$ 170km/s (where $V_{A(pol)} = B_{in(z)}/\sqrt{\mu_0 \rho}$). 
The continuous spontaneous breaking of the current sheet and the subsequent formation of current-carrying magnetically self-confined loops, i.e. plasmoids, occurs at later times (Fig.~\ref{fig:fig4}(b-c)), as long as the voltage is applied. As seen from  the Poincare plot in Fig.~\ref{fig:fig4}(c), two large plasmoids are formed. At this later time, as the first plasmoid is already ejected, the open field lines start to close again and a large-volume closed field lines in the form of a second plasmoid is formed and departs the device with a high outflow velocity of about 460 km/s (Fig.~\ref{fig:fig4}(b). We should note that small-scale plasmoids are also formed along a current sheet (seen in red color in Fig.~\ref{fig:fig3}(c)) above the primary exhaust current sheet. Although these also contribute to  a positive large vertical outflows, they are less important, as they are not directly connected to the injector plates. It is important to note that although all the simulations above are 3-D (by including 21 non-axisymmetric ($n_{\phi}\neq 0$) modes), all the structures during nonlinear evolution shown in Figs. 2-4 remain azimuthally symmetric (axisymmetric). In these simulations,  the time scale for the cyclic ejection of large scale axisymmetric plasmoids is around 10 $\mu$s (4-5 Alfven transit times). This time-scale is much shorter for any nonaxisymmetic disturbance to growth to large amplitudes.

\begin{figure}
\centering
    \includegraphics[width=3.5in,height=2.5in]{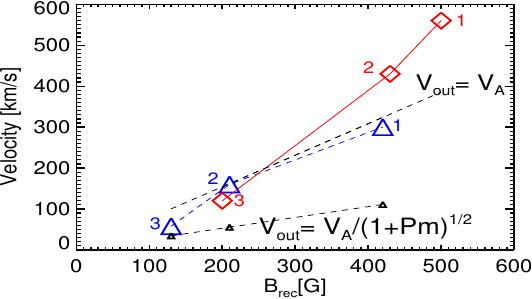}
    \caption{Maximum exhaust velocity obtained from simulations vs. reconnecting magnetic field for two magnetic configurations with different ratio of poloidal coil currents. 
     The dashed black line with triangles shows the theoretical S-P scaling.}
  \label{fig:fig5} 
\end{figure}

\section{Scaling of the exhaust velocity with reconnecting field}
To further examine the variation and the dependence of the exhaust velocity on the injected field ($B_z$), we have performed simulations with two different magnetic configurations in which we have varied the coil currents.  We should note that in all the simulations presented above the azimuthal field is intrinsically generated.  Here we also use an external azimuthal guide field of about 7000G, which would not affect the scaling, as we only use the reconnecting $B_z$ field in Fig.~\ref{fig:fig5}.  The first configuration is shown in Fig.~\ref{fig:fig5}
in blue triangles at $B_{\mathrm{rec}}$ =400G (case 1), which results in  $V_z$ =300 km/s. To scan over $B_{\mathrm{rec}}$,  we first reduce the detachment coil current (which gives $B_z$= 200G, case 2), and then we reduce the current in all the coils by 50 percent (gives $B_z$= 130G, case 3), and much reduced outflow velocity of about 60km/s. In the second configuration, we use a much higher
ratio of the two shaping coil currents, but with the same injector and detachment coil currents as the first configuration. The dependency of the outflow on $B_{\mathrm{rec}}$ for the cases in this configuration is also shown (by red diamonds) in Fig.~\ref{fig:fig5}.   We also  compare the outflows (maximum flows along the exhaust current sheet) obtained in the simulations with the slow S-P type reconnection model outflows in a visco-resistive MHD plasma (with Pm=7.5). Using $B_z$ from the simulations, Fig.~\ref{fig:fig5} shows the calculated outflows based on the local S-P model (black triangles) vs. $B_z$. 
It is shown that the outflows from MHD simulations 1) do favorably scale with $B_z$, 2) are higher (more than  double) compared to the slow S-P velocities, and are due to the fast plasmoid-mediated reconnection. It is therefore found that the exhaust velocities in these magnetic configurations are Alfvenic-type reconnecting outflows as they are strongly
associated with the strength of  $B_{\mathrm{rec}}$. 

\section{Thrust and the thrust to power ratio}

Because the plasmoids are ejected at the Alfven velocity, the expression for the thrust~\citep{lovberg197116} becomes $F = \rho \, V_A^2 \, A$
, where $A$ is the area of the plasmoid cross section. Notably, the thrust does then not depend on $\rho$, and it scales as the magnetic field squared ($B^2$). For example, for plasmoids with radius 10cm (as in Fig. 3(d)) and reconnecting field of B=800G, the calculated thrust is about 50N, taking into account a duty cycle of about 33\% (i.e. the distance between two consecutive plasmoids is twice the plasmoid length). The input power is given by $P_{\mathrm{inj}}$= $I_{\mathrm{inj}} \, V_{\mathrm{inj}}$, where $I_{\mathrm{inj}}$ = $ 2\pi rB_{\phi}/\mu_0 $.  In general $I_{\mathrm{inj}}$ could vary from a few to a few hundred kA. In our simulations,  $I_{\mathrm{inj}}$ is about 100 kA (equivalent to $B_{\phi}\sim$ 500G), corresponding to about 10 MW of power. For this unoptimized high-power case (with a trust of 50-100N), the ratio of thrust over power is thus about 5-10 mN/kW. We have not yet performed a systematic optimization, but tentatively the optimal parameter range for this new thruster will be $I_{SP}$ (specific impulse) from 2,000 to 50,000 s, power from 0.1 to 10 MW and thrust from 1 to 100 Newtons. It would thus occupy a complementary part of parameter space with little overlap with existing thrusters.

In helicity injection startup plasma
experiments (with an injection region similar to here), plasma has been efficiently produced, and both plasma and magnetic fields have been successfully injected via an injector gap~\citep{raman2003,raman2011experimental}.
The fundamentals of plasma production and ionization for this concept are
essentially the same as for an unmagnetized DC gas discharge. As shown by~\citep{hammond2017}, for keeping the operating voltage in a reasonable range of a few hundred volts (for acceptable cathode sputtering and good ionization efficiency), the Paschen curve imposes a minimum gas pressure. For example, for our application the connection length ($L_c$) is about 10~cm (depending on the vertical and azimuthal magnetic fields), which requires a gas pressure of tens of mTorrs (we used $L_c P$ of about 6 Torr x mm given by ~\citep{hammond2017}, for an operating point reasonably close to the Paschen minimum).
Operating voltages from a  few hundred up to a thousand volts have routinely been used for helicity injection experiments, including plasma accelerators as well as plasma startup for current-drive. Significant cathode erosion (from sputtering or arcing) in the injector region has not been reported. For long-pulse operation, the cathode is sometimes coated with graphite or tungsten to minimize sputtering [in Refs. \citep{raman2003} and in other helicity injection experiments].
Once the plasmoid has formed, the simulations show that it stays away from the walls and should therefore not contribute to wall erosion. In the  simulations walls provide the necessary boundary conditions in
the domain, however more
evolved versions of this thruster might in fact be wall less. The details of neutral dynamics also remain for future work.



%
\section{Summary}
Here, we have presented a new concept for generation of thrust for space propulsion. With a low plasma temperature of only a few eV, the plasmoid objects, which could have diameters as large as several tens of centimeters, are generated in a fluid-like (MHD and two-fluid Hall) regime and move with the center of mass of plasma. 
The concept is explored via 3-D extended MHD simulations of reconnecting plasmoid formation during helicity injection into an annular channel. Based on the simulations above, we find that there are fundamentally several advantages of this novel thruster, including: 1- High and variable exhaust velocity as large as 500km/s with injected poloidal field of 500-600G. 2- Large and scalable thrust --  depending on the size of plasmoid and magnetic field strength, the thrust can range at least from a tenth of a Newton to tens of Newtons. As the reconnecting plasmoids leave the device at the Alfven velocity, the thrust scales as magnetic field squared. 3- The thrust does not ideally depend on ion mass, so plasma can be created from a wide range of gases, including gases extracted from asteroids. We should note that
reconnection process is advantageous for space propulsion, as the detachment from the magnetic field in the nozzle~\citep{detachment} is not an issue here. Plasmoids
are closed magnetic structures, they are detached from the moment they are
created.

Lastly, the experimental NSTX camera images during helicity injection plasma startup in ~\citep{ebrahimi2015plasmoids} (and the supplementary movie there), which show distinct plasmoids leaving the device with velocities of about 25km/s, have inspired this thruster concept and could in fact provide a proof of principle. The first qualitative experimental evidence of plasmoid formation demonstrated there was first predicted by global MHD simulations~\citep{ebrahimi2015plasmoids}, later expanded for plasmoid-driven startup in spherical tokamaks ~\citep{ebrahimi16,ebrahimi2019}.
The extended MHD simulations presented here have been instrumental for exploring the fundamental physics of this new concept. However, more detailed physics (for example neutral dynamics and multi-fluid effects) could be numerically investigated in a future study to develop predictive capabilities for building a prototype device.

\section{Acknowledgements}
The author acknowledges insightful comments by Jon Menard and Stewart Prager. Computations were performed at NERSC and local PPPL cluster. This work was supported by DOE grants DE-AC02-09CHI1466, and DE-SC0010565; and PPPL LDRD grant.


\begin{thebibliography}{51}
\expandafter\ifx\csname natexlab\endcsname\relax\def\natexlab#1{#1}\fi
\def\au#1{#1} \def\ed#1{#1} \def\yr#1{#1}\def\at#1{#1}\def\jt#1{\textit{#1}}
  \def\bt#1{#1}\def\bvol#1{\textbf{#1}} \def\vol#1{#1} \def\pg#1{#1}
  \def\publ#1{#1}\def\arxiv#1{#1}\def\org#1{#1}\def\st#1{\textit{#1}}

\bibitem[{Alfven} {\em et~al.\/}(1960){Alfven}, {Lindberg} \&
  {Mitlid}]{alfven60}
{\sc \au{{Alfven}, H.}, \au{{Lindberg}, L.} \& \au{{Mitlid}, P.}} \yr{1960}
  \at{{Experiments with plasma rings}}.  \jt{Journal of Nuclear Energy}
  \bvol{1},  \pg{116--120}.

\bibitem[Arefiev \& Breizman(2005)]{detachment}
{\sc \au{Arefiev, Alexey~V} \& \au{Breizman, Boris~N}} \yr{2005}
  \at{Magnetohydrodynamic scenario of plasma detachment in a magnetic nozzle}.
  \jt{Physics of Plasmas}  \bvol{12}~(4),  \pg{043504}.

\bibitem[Bathgate {\em et~al.\/}(2018)Bathgate, Bilek, Cairns \&
  McKenzie]{bathgate2018thruster}
{\sc \au{Bathgate, Stephen~N}, \au{Bilek, Marcela~MM}, \au{Cairns, Iver~H} \&
  \au{McKenzie, David~R}} \yr{2018}  \at{A thruster using magnetic reconnection
  to create a high-speed plasma jet.}  \jt{European Physical Journal-Applied
  Physics}  \bvol{84}~(2).

\bibitem[{Bayliss} {\em et~al.\/}(2011){Bayliss}, {Sovinec} \& {Redd}]{adamchi}
{\sc \au{{Bayliss}, R.~A.}, \au{{Sovinec}, C.~R.} \& \au{{Redd}, A.~J.}}
  \yr{2011}  \at{{Zero-{$\beta$} modeling of coaxial helicity injection in the
  HIT-II spherical torus}}.  \jt{{Phys. Plasmas}}  \bvol{18}~(9),  \pg{094502}.

\bibitem[{Biskamp}(1986)]{biskamp86}
{\sc \au{{Biskamp}, D.}} \yr{1986}  \at{{Magnetic reconnection via current
  sheets}}.  \jt{Physics of Fluids}  \bvol{29},  \pg{1520--1531}.

\bibitem[{Brown} \& {Bellan}(1990)]{brown90}
{\sc \au{{Brown}, M.~R.} \& \au{{Bellan}, P.~M.}} \yr{1990}  \at{{Current drive
  by spheromak injection into a tokamak}}.  \jt{Physical Review Letters}
  \bvol{64},  \pg{2144--2147}.

\bibitem[{Cassak} {\em et~al.\/}(2005){Cassak}, {Shay} \& {Drake}]{cassak2005}
{\sc \au{{Cassak}, P.~A.}, \au{{Shay}, M.~A.} \& \au{{Drake}, J.~F.}} \yr{2005}
   \at{{Catastrophe Model for Fast Magnetic Reconnection Onset}}.  \jt{Physical
  Review Letters}  \bvol{95}~(23),  \pg{235002},  \arxiv{arXiv:
  physics/0502001}.

\bibitem[{Cheng}(1971)]{cheng_gun71}
{\sc \au{{Cheng}, D.~Y.}} \yr{1971}  \at{{Application of a deflagration plasma
  gun as a space propulsion thruster}}.  \jt{AIAA Journal}  \bvol{9},
  \pg{1681--1685}.

\bibitem[{Chesny} {\em et~al.\/}(2017){Chesny}, {Orange}, {Oluseyi} \&
  {Valletta}]{2017JPlPh}
{\sc \au{{Chesny}, David~L.}, \au{{Orange}, N.~Brice}, \au{{Oluseyi},
  Hakeem~M.} \& \au{{Valletta}, David~R.}} \yr{2017}  \at{{Toward laboratory
  torsional spine magnetic reconnection}}.  \jt{Journal of Plasma Physics}
  \bvol{83}~(6),  \pg{905830602}.

\bibitem[Choueiri(2009)]{Choueiri2009}
{\sc \au{Choueiri, Edgar~Y}} \yr{2009}  \at{New dawn for electric rockets}.
  \jt{Scientific American}  \bvol{300}~(2),  \pg{58--65}.

\bibitem[Comisso {\em et~al.\/}(2016)Comisso, Lingam, Huang \&
  Bhattacharjee]{2016luca}
{\sc \au{Comisso, L}, \au{Lingam, M}, \au{Huang, Y-M} \& \au{Bhattacharjee, A}}
  \yr{2016}  \at{General theory of the plasmoid instability}.  \jt{Physics of
  Plasmas}  \bvol{23},  \pg{100702}.

\bibitem[Dale {\em et~al.\/}(2020)Dale, Jorns \& Gallimore]{dale2020future}
{\sc \au{Dale, Ethan}, \au{Jorns, Benjamin} \& \au{Gallimore, Alec}} \yr{2020}
  \at{Future directions for electric propulsion research}.  \jt{Aerospace}
  \bvol{7}~(9),  \pg{120}.

\bibitem[{Daughton} {\em et~al.\/}(2009){Daughton}, {Roytershteyn}, {Albright},
  {Karimabadi}, {Yin} \& {Bowers}]{daugthon09}
{\sc \au{{Daughton}, W.}, \au{{Roytershteyn}, V.}, \au{{Albright}, B.~J.},
  \au{{Karimabadi}, H.}, \au{{Yin}, L.} \& \au{{Bowers}, K.~J.}} \yr{2009}
  \at{{Transition from collisional to kinetic regimes in large-scale
  reconnection layers}}.  \jt{Physical Review Letters}  \bvol{103}~(6),
  \pg{065004}.

\bibitem[Ebrahimi(2016)]{ebrahimi2016dynamo}
{\sc \au{Ebrahimi, F}} \yr{2016}  \at{Dynamo-driven plasmoid formation from a
  current-sheet instability}.  \jt{Physics of Plasmas}  \bvol{23}~(12).

\bibitem[Ebrahimi(2017)]{ebrahimi2017ELM}
{\sc \au{Ebrahimi, F}} \yr{2017}  \at{Nonlinear reconnecting edge localized
  modes in current-carrying plasmas}.  \jt{Physics of Plasmas}  \bvol{24}~(5),
  \pg{056119}.

\bibitem[Ebrahimi(2019)]{ebrahimi2019}
{\sc \au{Ebrahimi, F}} \yr{2019}  \at{Three-dimensional plasmoid-mediated
  reconnection and the effect of toroidal guide field in simulations of coaxial
  helicity injection}.  \jt{Physics of Plasmas}  \bvol{26}~(9),  \pg{092502}.

\bibitem[Ebrahimi {\em et~al.\/}(2013)Ebrahimi, Hooper, Sovinec \&
  Raman]{ebrahimi2013}
{\sc \au{Ebrahimi, F.}, \au{Hooper, E.~B.}, \au{Sovinec, C.~R.} \& \au{Raman,
  R.}} \yr{2013}  \at{Magnetic reconnection process in transient coaxial
  helicity injection}.  \jt{Physics of Plasmas}  \bvol{20},  \pg{090702}.

\bibitem[Ebrahimi \& Prager(2011)]{ebrahimitearing}
{\sc \au{Ebrahimi, F.} \& \au{Prager, S.~C.}} \yr{2011}  \at{Momentum tranport
  from current-driven reconnection in astrophysical disks}.  \jt{Astrophys. J.}
   \bvol{740},  \pg{http://arxiv.org/abs/1109.5763}.

\bibitem[Ebrahimi \& Raman(2015)]{ebrahimi2015plasmoids}
{\sc \au{Ebrahimi, F} \& \au{Raman, R}} \yr{2015}  \at{Plasmoids formation
  during simulations of coaxial helicity injection in the national spherical
  torus experiment}.  \jt{Physical Review Letters}  \bvol{114}~(20),
  \pg{205003}.

\bibitem[{Ebrahimi} \& {Raman}(2016)]{ebrahimi16}
{\sc \au{{Ebrahimi}, F.} \& \au{{Raman}, R.}} \yr{2016}  \at{{Large-volume flux
  closure during plasmoid-mediated reconnection in coaxial helicity
  injection}}.  \jt{Nuclear Fusion}  \bvol{56}~(4),  \pg{044002}.

\bibitem[Fox {\em et~al.\/}(2016)Fox, Velli, Bale, Decker, Driesman, Howard,
  Kasper, Kinnison, Kusterer, Lario {\em et~al.\/}]{fox2016PSP}
{\sc \au{Fox, NJ}, \au{Velli, MC}, \au{Bale, SD}, \au{Decker, R}, \au{Driesman,
  A}, \au{Howard, RA}, \au{Kasper, Justin~C}, \au{Kinnison, J}, \au{Kusterer,
  M}, \au{Lario, D} \& \au{others}} \yr{2016}  \at{The solar probe plus
  mission: humanity’s first visit to our star}.  \jt{Space Science Reviews}
  \bvol{204}~(1-4),  \pg{7--48}.

\bibitem[Goebel \& Katz(2008)]{goebel2008}
{\sc \au{Goebel, Dan~M} \& \au{Katz, Ira}} \yr{2008} {\em Fundamentals of
  electric propulsion: ion and Hall thrusters\/}, ,  \vol{vol.~1}.  \publ{John
  Wiley \& Sons}.

\bibitem[G\"unter {\em et~al.\/}(2015)G\"unter, Yu, Lackner, Bhattacharjee \&
  Huang]{gunter15}
{\sc \au{G\"unter, S}, \au{Yu, Q}, \au{Lackner, K}, \au{Bhattacharjee, A} \&
  \au{Huang, Y-M}} \at{ \yr{2015} } \jt{Plasma Physics and Controlled Fusion}
  \bvol{57},  \pg{014017}.

\bibitem[{Hammer} {\em et~al.\/}(1988){Hammer}, {Hartman}, {Eddleman} \&
  {McLean}]{ring_CT88}
{\sc \au{{Hammer}, J.~H.}, \au{{Hartman}, C.~W.}, \au{{Eddleman}, J.~L.} \&
  \au{{McLean}, H.~S.}} \yr{1988}  \at{{Experimental demonstration of
  acceleration and focusing of magnetically confined plasma rings}}.
  \jt{Physical Review Letters}  \bvol{61},  \pg{2843--2846}.

\bibitem[Hammond {\em et~al.\/}(2017)Hammond, Raman \& Volpe]{hammond2017}
{\sc \au{Hammond, KC}, \au{Raman, R} \& \au{Volpe, FA}} \yr{2017}
  \at{Application of townsend avalanche theory to tokamak startup by coaxial
  helicity injection}.  \jt{Nuclear Fusion}  \bvol{58}~(1),  \pg{016013}.

\bibitem[Hooper {\em et~al.\/}(2012)Hooper, Bulmer, Cohen, Hill, Holcomb,
  Hudson, McLean, Pearlstein, Romero-Talam{\'a}s, Sovinec {\em
  et~al.\/}]{hooper2012}
{\sc \au{Hooper, E~Biclcford}, \au{Bulmer, RH}, \au{Cohen, BI}, \au{Hill, DN},
  \au{Holcomb, CT}, \au{Hudson, B}, \au{McLean, HS}, \au{Pearlstein, LD},
  \au{Romero-Talam{\'a}s, CA}, \au{Sovinec, CR} \& \au{others}} \yr{2012}
  \at{Sustained spheromak physics experiment (sspx): design and physics
  results}.  \jt{Plasma Physics and Controlled Fusion}  \bvol{54}~(11),
  \pg{113001}.

\bibitem[Hooper {\em et~al.\/}(2013)Hooper, Sovinec, Raman, Ebrahimi \&
  Menard]{hooper2013}
{\sc \au{Hooper, E.~B.}, \au{Sovinec, C.~R.}, \au{Raman, R.}, \au{Ebrahimi, F.}
  \& \au{Menard, J.~E.}} \yr{2013}  \at{Resistive magnetohydrodynamic
  simulations of helicity-injected startup plasmas in national spherical torus
  experiment}.  \jt{Physics of Plasmas}  \bvol{20},  \pg{092510}.

\bibitem[Hsu \& Bellan(2003)]{hsu2003}
{\sc \au{Hsu, Scott~C} \& \au{Bellan, Paul~M}} \yr{2003}  \at{Experimental
  identification of the kink instability as a poloidal flux amplification
  mechanism for coaxial gun spheromak formation}.  \jt{Physical review letters}
   \bvol{90}~(21),  \pg{215002}.

\bibitem[Huang {\em et~al.\/}(2011)Huang, Bhattacharjee \&
  Sullivan]{huang2011hall}
{\sc \au{Huang, Yi-Min}, \au{Bhattacharjee, A} \& \au{Sullivan, Brian~P}}
  \yr{2011}  \at{Onset of fast reconnection in hall magnetohydrodynamics
  mediated by the plasmoid instability}.  \jt{Physics of Plasmas}
  \bvol{18}~(7),  \pg{072109--072109}.

\bibitem[{Jarboe} {\em et~al.\/}(1983){Jarboe}, {Henins}, {Sherwood}, {Barnes}
  \& {Hoida}]{jarboe83}
{\sc \au{{Jarboe}, T.~R.}, \au{{Henins}, I.}, \au{{Sherwood}, A.~R.},
  \au{{Barnes}, C.~W.} \& \au{{Hoida}, H.~W.}} \yr{1983}  \at{{Slow Formation
  and Sustainment of Spheromaks by a Coaxial Magnetized Plasma Source}}.
  \jt{Physical Review Letters}  \bvol{51},  \pg{39--42}.

\bibitem[Ji {\em et~al.\/}(2020)Ji, Alt, Antiochos, Baalrud, Bale, Bellan,
  Begelman, Beresnyak, Blackman, Brennan {\em et~al.\/}]{ji2020major}
{\sc \au{Ji, Hantao}, \au{Alt, A}, \au{Antiochos, S}, \au{Baalrud, S},
  \au{Bale, S}, \au{Bellan, PM}, \au{Begelman, M}, \au{Beresnyak, A},
  \au{Blackman, EG}, \au{Brennan, D} \& \au{others}} \yr{2020}  \at{Major
  scientific challenges and opportunities in understanding magnetic
  reconnection and related explosive phenomena throughout the universe}.
  \jt{arXiv preprint arXiv:2004.00079} .

\bibitem[{Ji} \& {Daughton}(2011)]{ji2011}
{\sc \au{{Ji}, H.} \& \au{{Daughton}, W.}} \yr{2011}  \at{{Phase diagram for
  magnetic reconnection in heliophysical, astrophysical, and laboratory
  plasmas}}.  \jt{Physics of Plasmas}  \bvol{18}~(11),  \pg{111207},
  \arxiv{arXiv: 1109.0756}.

\bibitem[{Loureiro} {\em et~al.\/}(2007){Loureiro}, {Schekochihin} \&
  {Cowley}]{loureiro07}
{\sc \au{{Loureiro}, N.~F.}, \au{{Schekochihin}, A.~A.} \& \au{{Cowley},
  S.~C.}} \yr{2007}  \at{{Instability of current sheets and formation of
  plasmoid chains}}.  \jt{Physics of Plasmas}  \bvol{14}~(10),  \pg{100703},
  \arxiv{arXiv: astro-ph/0703631}.

\bibitem[Lovberg(1971)]{lovberg197116}
{\sc \au{Lovberg, Ralph~H}} \yr{1971}  \at{16. plasma problems in electrical
  propulsion}.  \bt{In {\em Methods in Experimental Physics\/}}, ,
  \vol{vol.~9},  \pg{pp. 251--289}.  \publ{Elsevier}.

\bibitem[{Marshall}(1960)]{marshall1960}
{\sc \au{{Marshall}, J.}} \yr{1960}  \at{{Performance of a Hydromagnetic Plasma
  Gun}}.  \jt{Physics of Fluids}  \bvol{3},  \pg{134--135}.

\bibitem[McLean {\em et~al.\/}(2002)McLean, Woodruff, Hooper, Bulmer, Hill,
  Holcomb, Moller, Stallard, Wood \& Wang]{mclean2002}
{\sc \au{McLean, HS}, \au{Woodruff, S}, \au{Hooper, EB}, \au{Bulmer, RH},
  \au{Hill, DN}, \au{Holcomb, C}, \au{Moller, J}, \au{Stallard, BW}, \au{Wood,
  RD} \& \au{Wang, Z}} \yr{2002}  \at{Suppression of mhd fluctuations leading
  to improved confinement in a gun-driven spheromak}.  \jt{Physical review
  letters}  \bvol{88}~(12),  \pg{125004}.

\bibitem[Morgan {\em et~al.\/}(2019)Morgan, Jarboe \&
  Akcay]{morgan2019formation}
{\sc \au{Morgan, Kyle}, \au{Jarboe, Thomas} \& \au{Akcay, Cihan}} \yr{2019}
  \at{Formation of closed flux surfaces in spheromaks sustained by steady
  inductive helicity injection}.  \jt{Nuclear Fusion}  \bvol{59}~(6),
  \pg{066037}.

\bibitem[Morozov {\em et~al.\/}(1972)Morozov, Esipchuk, Kapulkin, Nevrovskii \&
  Smirnov]{Morozov1972}
{\sc \au{Morozov, AI}, \au{Esipchuk, Yu~V}, \au{Kapulkin, AM}, \au{Nevrovskii,
  VA} \& \au{Smirnov, VA}} \yr{1972}  \at{Effect of the magnetic field a
  closed-electron-drift accelerator}.  \jt{Soviet Physics Technical Physics}
  \bvol{17},  \pg{482}.

\bibitem[Raitses {\em et~al.\/}(2007)Raitses, Smirnov \& Fisch]{Raitses2007}
{\sc \au{Raitses, Y}, \au{Smirnov, A} \& \au{Fisch, Nathaniel~J}} \yr{2007}
  \at{Enhanced performance of cylindrical hall thrusters}.  \jt{Applied physics
  letters}  \bvol{90}~(22),  \pg{221502}.

\bibitem[Raman(2020)]{private}
{\sc \au{Raman, R.}} \yr{2020} private communication.

\bibitem[{Raman} {\em et~al.\/}(2003){Raman}, {Jarboe}, {Nelson}, {Izzo},
  {O'Neill}, {Redd} \& {Smith}]{raman2003}
{\sc \au{{Raman}, R.}, \au{{Jarboe}, T.~R.}, \au{{Nelson}, B.~A.}, \au{{Izzo},
  V.~A.}, \au{{O'Neill}, R.~G.}, \au{{Redd}, A.~J.} \& \au{{Smith}, R.~J.}}
  \yr{2003}  \at{{Demonstration of Plasma Startup by Coaxial Helicity
  Injection}}.  \jt{Phys. Rev. Lett.}  \bvol{90}~(7),  \pg{075005--+}.

\bibitem[{Raman} {\em et~al.\/}(1994){Raman}, {Martin}, {Quirion}, {St-Onge},
  {Lachambre}, {Michaud}, {Sawatzky}, {Thomas}, {Hirose}, {Hwang}, {Richard},
  {C{\^o}t{\'e}}, {Abel}, {Pinsonneault}, {Gauvreau}, {Stansfield},
  {D{\'e}coste}, {C{\^o}t{\'e}}, {Zuzak} \& {Boucher}]{raman_CT94}
{\sc \au{{Raman}, R.}, \au{{Martin}, F.}, \au{{Quirion}, B.}, \au{{St-Onge},
  M.}, \au{{Lachambre}, J.-L.}, \au{{Michaud}, D.}, \au{{Sawatzky}, B.},
  \au{{Thomas}, J.}, \au{{Hirose}, A.}, \au{{Hwang}, D.}, \au{{Richard}, N.},
  \au{{C{\^o}t{\'e}}, C.}, \au{{Abel}, G.}, \au{{Pinsonneault}, D.},
  \au{{Gauvreau}, J.-L.}, \au{{Stansfield}, B.}, \au{{D{\'e}coste}, R.},
  \au{{C{\^o}t{\'e}}, A.}, \au{{Zuzak}, W.} \& \au{{Boucher}, C.}} \yr{1994}
  \at{{Experimental demonstration of nondisruptive, central fueling of a
  tokamak by compact toroid injection}}.  \jt{Physical Review Letters}
  \bvol{73},  \pg{3101--3104}.

\bibitem[Raman {\em et~al.\/}(2011)Raman, Mueller, Jarboe, Nelson, Bell,
  Gerhardt, LeBlanc, Menard, Ono, Roquemore {\em
  et~al.\/}]{raman2011experimental}
{\sc \au{Raman, R}, \au{Mueller, D}, \au{Jarboe, TR}, \au{Nelson, BA},
  \au{Bell, MG}, \au{Gerhardt, S}, \au{LeBlanc, B}, \au{Menard, J}, \au{Ono,
  M}, \au{Roquemore, L} \& \au{others}} \yr{2011}  \at{Experimental
  demonstration of tokamak inductive flux saving by transient coaxial helicity
  injection on national spherical torus experiment}.  \jt{Physics of Plasmas}
  \bvol{18}~(9),  \pg{092504}.

\bibitem[Ripperda {\em et~al.\/}(2020)Ripperda, Bacchini \&
  Philippov]{ripperda2020magnetic}
{\sc \au{Ripperda, Bart}, \au{Bacchini, Fabio} \& \au{Philippov, Alexander}}
  \yr{2020}  \at{Magnetic reconnection and hot-spot formation in black-hole
  accretion disks}.  \jt{arXiv preprint arXiv:2003.04330} .

\bibitem[{Schoenberg} {\em et~al.\/}(1993){Schoenberg}, {Gerwin}, {Henins},
  {Mayo}, {Scheuer} \& {Wurden}]{Schoenberg93}
{\sc \au{{Schoenberg}, K.~F.}, \au{{Gerwin}, R.~A.}, \au{{Henins}, I.},
  \au{{Mayo}, R.~M.}, \au{{Scheuer}, J.~T.} \& \au{{Wurden}, G.~A.}} \yr{1993}
  \at{{Preliminary investigation of power flow and performance phenomena in a
  multimegawatt coaxial plasma thruster}}.  \jt{IEEE Transactions on Plasma
  Science}  \bvol{21},  \pg{625--644}.

\bibitem[Sovinec {\em et~al.\/}(2004)Sovinec, Glasser, Gianakon, Barnes, Nebel,
  Kruger, Schnack, Plimpton, Tarditi, Chu \& the NIMROD~Team]{sovinec04}
{\sc \au{Sovinec, C.~R.}, \au{Glasser, A.~H.}, \au{Gianakon, T.~A.},
  \au{Barnes, D.~C.}, \au{Nebel, R.~A.}, \au{Kruger, S.~E.}, \au{Schnack,
  D.~D.}, \au{Plimpton, S.~J.}, \au{Tarditi, A.}, \au{Chu, M.} \& \au{the
  NIMROD~Team}} \yr{2004}  \at{Nonlinear magnetohydrodynamics simulation using
  high-order finite elements.}  \jt{J. Comput. Phys.}  \bvol{195},  \pg{355}.

\bibitem[Stuhlinger(1964)]{Stuhlinger1964}
{\sc \au{Stuhlinger, Ernst}} \yr{1964} {\em Ion propulsion for space flight\/}.
   \publ{McGraw-Hill New York}.

\bibitem[{Tajima} \& {Shibata}(1997)]{tajima_shibata97}
{\sc \au{{Tajima}, T.} \& \au{{Shibata}, K.}}, ed. \yr{1997} {\em {Plasma
  astrophysics}\/}.

\bibitem[{Witherspoon} {\em et~al.\/}(2009){Witherspoon}, {Case}, {Messer},
  {Bomgardner}, {Phillips}, {Brockington} \& {Elton}]{witherspoon2009}
{\sc \au{{Witherspoon}, F.~D.}, \au{{Case}, A.}, \au{{Messer}, S.~J.},
  \au{{Bomgardner}, R.}, \au{{Phillips}, M.~W.}, \au{{Brockington}, S.} \&
  \au{{Elton}, R.}} \yr{2009}  \at{{A contoured gap coaxial plasma gun with
  injected plasma armature}}.  \jt{Review of Scientific Instruments}
  \bvol{80}~(8),  \pg{083506--083506--15}.

\bibitem[Yoo {\em et~al.\/}(2017)Yoo, Na, Jara-Almonte, Yamada, Ji,
  Roytershteyn, Argall, Fox \& Chen]{yoo2017electron}
{\sc \au{Yoo, Jongsoo}, \au{Na, Byungkeun}, \au{Jara-Almonte, Jonathan},
  \au{Yamada, Masaaki}, \au{Ji, Hantao}, \au{Roytershteyn, Vadim}, \au{Argall,
  Matthew~R}, \au{Fox, William} \& \au{Chen, Li-Jen}} \yr{2017}  \at{Electron
  heating and energy inventory during asymmetric reconnection in a laboratory
  plasma}.  \jt{Journal of Geophysical Research: Space Physics}
  \bvol{122}~(9),  \pg{9264--9281}.

\bibitem[{Zweibel} \& {Yamada}(2009)]{zweibel09}
{\sc \au{{Zweibel}, E.~G.} \& \au{{Yamada}, M.}} \yr{2009}  \at{{Magnetic
  Reconnection in Astrophysical and Laboratory Plasmas}}.  \jt{Annual Review of
  Astronomy and Astrophysics}  \bvol{47},  \pg{291--332}.

\end{thebibliography}

\end{document}